\documentclass[final,11pt, letterpaper]{article}
\usepackage[noblocks]{authblk}

\usepackage{amsmath}
\usepackage{array}
\usepackage{booktabs}
\usepackage{xcolor}
\usepackage[inline]{enumitem}
\usepackage[pdftex,final]{graphicx}
\usepackage{hyperref}
\usepackage[labelfont=bf,
            small,
            labelsep=period,
            figurename=Figure]
            {caption}
\usepackage{setspace}
\usepackage[mathlines]{lineno}
\usepackage[compact,sf,bf]{titlesec}

\usepackage[letterpaper,margin=1in]{geometry}
\usepackage[authoryear,sort&compress,comma]{natbib}
\graphicspath{{}}

\def\fig{figure}

\date{}

\title{\Large Transverse contributions to the longitudinal stiffness of the human foot} 

\author[1]{Ali Yawar\thanks{ali.yawar@aya.yale.edu}}
\author[1]{Lucia Korpas}
\author[2]{Shreyas Mandre}
\author[1]{Madhusudhan Venkadesan\thanks{m.venkadesan@yale.edu}}

\affil[1]{\footnotesize Department of Mechanical Engineering \& Materials Science, Yale University, New Haven, CT, USA}
\affil[2]{\footnotesize Department of Engineering, University of Cambridge, Cambridge, UK}

\begin{document}

\maketitle

\section*{Abstract}
Humans rely on foot stiffness to withstand the propulsive forces of walking and running.
Skeletal adaptations that increase foot stiffness include the medial longitudinal arch (MLA) and the transverse tarsal arch (TTA).
The TTA has been hypothesized to stiffen the foot through cross-axis coupling of transverse intermetatarsal stiffness with sagittal-plane midfoot stiffness, but this has been tested only in cadaveric specimens.
{\itshape{}In vivo\/} testing is essential because muscle contraction substantially modulates MLA function and may similarly affect the TTA's cross-axis coupling.
Here we provide {\itshape{}in vivo\/} evidence for the TTA's contribution to foot stiffness by externally increasing intermetatarsal stiffness and measuring its effects on midfoot elasticity during walking. 
As predicted by the cross-axis coupling hypothesis, increasing intermetatarsal stiffness with an elastic tape wrapped around the forefoot reduced the energy absorbed in midfoot flattening and increased sagittal-plane midfoot stiffness concomitantly (mean,$\pm$,standard error of the mean (SEM): $13.9\% \pm 3\%$ and $16.8\% \pm 5.8\%$, respectively).
However, taping did not change the curvature of the TTA, thereby isolating the effects of cross-axis coupling from morphological changes to the TTA.
Thus, forefoot taping modulates midfoot stiffness through cross-axis coupling and could provide a non-invasive means to manage pathological foot flexibility or enhance athletic performance.

\singlespacing{}

\noindent{\footnotesize\itshape{\bfseries Keywords:} midfoot stiffness, transverse arch, longitudinal arch, transverse stiffness, cross-axis coupling\/}

\noindent{\footnotesize\itshape{\bfseries Abbreviations:} MLA -- medial longitudinal arch, TTA -- transverse tarsal arch, SEM -- standard error of the mean\/}


\section{Introduction}

When humans walk and run, the midfoot experiences large sagittal plane torques that cause it to deform \citep{bruening2012analysis1}.
Foot stiffness resists this deformation and enables efficient propulsion using the forefoot \citep{Ray2020aa}.
The medial longitudinal arch (MLA) and the transverse tarsal arch (TTA) are key evolved features that increase midfoot stiffness \citep{holowka2018rethinking}.
The MLA stiffens the foot through a bow-and-string mechanism in the sagittal plane, engaging the plantar fascia and other longitudinal tissues \citep{ker1987}.
The TTA is hypothesized to create cross-axis coupling between transverse elastic tissue stiffness and longitudinal midfoot stiffness \citep{venkadesan2020stiffness}.
Consistent with this hypothesis, \citet{tang2025wrapping} found that wrapping an elastic tape in the transverse direction improved running economy, which they interpreted as arising from increased stiffness.
However, their manipulation of the transverse stiffness resulted in changes to both arches (increased arch height), and is not a clean test of the TTA's function alone.
The hypothesized TTA-driven cross-axis coupling has been demonstrated directly only in cadaveric feet and {\itshape{}in vivo\/} tests are needed to assess whether the effect persists when foot muscles are also actively engaged in stiffness modulation \citep{Mann1964PhasicActivityIntrinsica,kelly2014intrinsic,kelly2015active,welte2018influence,farris2019functional,holowka2020human}.

Cadaveric experiments help understand the baseline stiffness of the foot,
and show how passive soft tissue stiffness is transformed into longitudinal midfoot stiffness \citep{ker1987,huang1993biomechanical,venkadesan2020stiffness}.
Studying passive tissues also impacts our understanding of muscles that share similar lines of action \citep{kelly2014intrinsic,kelly2015active}.
But cadaveric studies cannot capture the complete picture.
Muscles have complex attachments that are not all parallel to other passive tissues and muscle contraction could substantially alter the contribution of the passive tissues.
Such modulation is known from {\itshape{}in vivo\/} measurements of MLA function in static loading \citep{kelly2012recruitment,kelly2014intrinsic,welte2018influence} and during locomotion \citep{Mann1964PhasicActivityIntrinsica,kelly2015active,farris2019functional,holowka2020human}.
However, in the absence of direct midfoot stiffness measurements under alterations to the TTA, we do not know to what extent the cross-axis coupling found in cadaveric feet applies when the muscles in the foot and leg are also actively engaged in load bearing and propulsion.

\begin{figure}[hbt]
  \centering\includegraphics[width=0.92\textwidth]{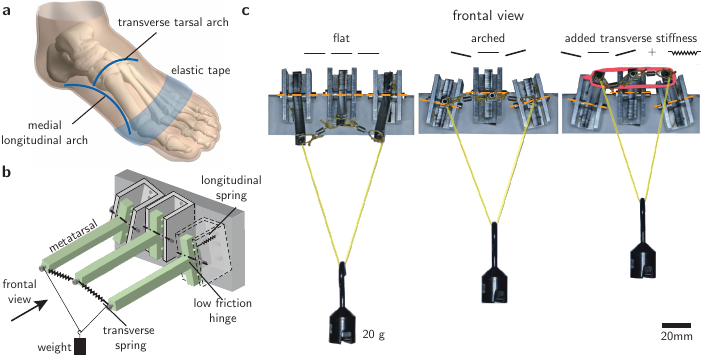}
  \caption[Cross-axis stiffness coupling hypothesis]{
    {\bfseries Cross-axis stiffness coupling hypothesis.}
    {\bfseries{a,}} Increasing the intermetatarsal stiffness using an elastic tape wound around the ball of the foot is predicted to increase the midfoot stiffness due to the TTA's cross-axis coupling.
    {\bfseries{b,}} Foot model with three hinged rods representing metatarsals.
    Transverse curvature implies non-parallel hinge axes (dash-dot lines). 
    The transverse curvature in the model is inverted with respect to the curvature in the foot so that gravity can be used as a substitute for ground reaction force at the distal end.
    {\bfseries{c,}} Frontal view of a physical foot model with a distal bending load. The model shows increasing longitudinal bending stiffness (left-to-right): without any transverse arch, with a transverse arch, with a transverse arch and a distal elastic band to increase the transverse intermetatarsal stiffness (highlighted in red).
  }%
\label{fig:intro}
\end{figure}

The mechanical principle for the hypothesized cross-axis coupling due to the TTA is similar to stiffening a pizza slice or a thin sheet of paper by slightly curling it.
Transverse curvature couples out of plane bending to in-plane stretching.
Trying to dorsiflex the midfoot in the sagittal plane engages the intermetatarsal tissues that resist splaying in the transverse direction and impart the foot with stiffness.
This was shown in cadaveric tests by surgically incising the transverse intermetatarsal tissues, whereupon the longitudinal foot stiffness decreased by over 40\% \citep{venkadesan2020stiffness}.
We cannot use such interventions {\itshape{}in vivo\/} to decrease the transverse stiffness, but we can increase it through external means like wrapping an elastic tape around the forefoot (\fig~\ref{fig:intro}a).

The cross-axis coupling can be demonstrated using a mechanical model of the mid and forefoot regions \citep[\fig~\ref{fig:intro}b,c]{venkadesan2020stiffness}.
The model has three metatarsals that are articulated by hinges at their base, and connected to the base and each other by longitudinal and transverse springs, respectively.
When the tip is loaded, the hinges rotate and stretch the longitudinal springs.
But when there is a transverse arch, the hinge axes are not parallel and cause the metatarsals to splay apart and engage the transverse springs.
The model with a transverse arch is stiffer than a flat model and deflects less under the same distal load (\fig~\ref{fig:intro}c).
To demonstrate the cross-axis coupling, we apply a transverse elastic band at the tip, whereupon the stiffness of the curved model increases even further (\fig~\ref{fig:intro}c).
This increase in stiffness is not due to an increase in the transverse curvature, which is identical before and after adding the elastic band (\fig~\ref{fig:intro}c).
Instead, stiffness increases because the elastic band further resists splaying of the model's metatarsals.
Similarly, adding transverse stiffness to the foot using an elastic band around the ball of the foot is predicted to increase the foot's stiffness and cause it to deform lesser under external loads, even if the arch geometry (curvature or height) is unaltered.
The distinction between changing TTA curvature (and height) and increasing transverse stiffness is important in light of other {\itshape{}in vivo\/} studies \citep{tang2025wrapping} that emphasized curvature changes but not stiffness modulation without changing foot shape.

We tested this prediction during walking in 13 consenting healthy human volunteers.
At the same walking speed, negative work associated with flattening of the midfoot is known to decrease when the foot is stiffer \citep{takahashi2016adding}. 
Thus, following established protocol \citep{Leardini2007,bruening2012analysis2,takahashi2017energy}, we measured how negative work at the midfoot, and the slope of the midfoot torque-angle curve changed upon tightly wrapping the forefoot with a stiff elastic tape.

\begin{figure}[tb!]
  \centering
  \includegraphics[width=0.5\textwidth]{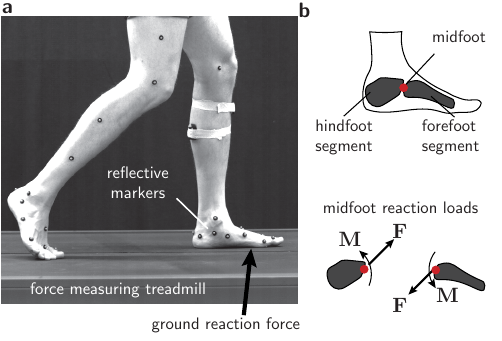}
  \caption[Experimental load-deformation measurements of the foot in walking.]{{\bfseries Experimental load-deformation measurements of the foot in walking.}
  {\bfseries a,} Subjects walked on a force-measuring treadmill with reflective markers placed on the left foot and leg according to the marker set described in \citep{Leardini2007}.
  {\bfseries b,} Bony landmarks on the foot define the hindfoot and forefoot segments.
  Rotational power is the dot product of the midfoot reaction torque with the angular velocity of the forefoot relative to the hindfoot.
  Translational power is the dot product of the midfoot joint reaction force with the translational velocity of the forefoot relative to the hindfoot.
  Their sum yields the six degree of freedom power of the midfoot \citep{zelik2015aa}.
  Assessment of repeatability of the tape application protocol is reported in \fig~\ref{fig:results:supplement:tape stiffness}, center of pressure trajectories in \fig~\ref{fig:results:supplement:COP medio-lateral traces} and \fig~\ref{fig:results:supplement:COP antero-posterior traces}, ankle angle in \fig~\ref{fig:results:supplement:ankle angle traces}. Motion capture accuracy estimates are reported in table~\ref{table:methods:supplement:calibration}.
  }\label{fig:methods}
\end{figure}

\section{Methods}

\subsection{Informed consent}
We performed human subject experiments with 13 healthy, consenting volunteers.
The cohort included 5 females and 8 males of age 29.7 years $\pm$ 4.8 years (mean\,$\pm$\,standard deviation).
Subject-wise age, weight, gender and foot size measurements are available in figure 3 -- source data 1.
Before the start of the experiment, the subjects studied the informed consent form and the experimenter discussed potential risks of the study and their option to withdraw from the study at any time.
The experiment was started after the subject agreed to participate in the study.
The detailed procedures for seeking informed consent were approved by Yale University's IRB.

\subsection{Data collection}
With the subject standing still on ground, we used calipers to measure the truncated foot length as the length from the posterior calcaneal tuberosity to the first metatarsophalangeal joint, and the forefoot width as the mediolateral distance between the distal heads of the first and fifth metatarsals.
Leg length is defined as the height of the greater trochanter above ground while standing in a neutral posture.
All subjects walked at the same non-dimensional speed $v = \sqrt{gL\,\rm Fr}$, where $g$ is the acceleration due to gravity, $L$ is leg length and the Froude number, $\rm Fr = 0.2$.
Reflective markers were placed on the lower limbs of the subjects following the Leardini marker set \citep{Leardini2007} (\fig~\ref{fig:methods}a).
After marker placement, subjects were instructed to walk on the treadmill for several minutes to allow them to acquaint themselves with the treadmill and its surroundings, and reach a steady walking cadence. 
Data recorded in the last 60 seconds of the walk are used in the analyses.
Three dimensional kinematics were recorded using marker-based optical motion capture at 500~Hz (Vicon Ltd.\ model T20S, Oxford, UK), and foot forces were recorded using a six-axis force plate at 2000~Hz (Bertec, Columbus, OH, USA).
Accuracy of the motion capture system was assessed by conducting dynamic measurements of marker distances and angles using a standard test object (5-marker calibration wand from Vicon).
Median and interquartile ranges for these measurements are reported in table~\ref{table:methods:supplement:calibration}.

The experiment was conducted under three conditions in a randomized order:
\begin{enumerate*}[label=(\textbf{\roman*})]
  \item free condition with no tape,
  \item taped condition with an elastic tape wrapped tightly around the distal metatarsal heads of the left foot, and
  \item control condition with the elastic tape loosely applied on the left foot with as little tension as possible.
\end{enumerate*}

We used commonly available adhesive Kinesiology tape (3B Scientific, Hamburg, Germany).
The length of the tape was customized for each subject to match twice the circumference of the unloaded forefoot.
In pilot trials, tape of lower length tended to peel off or tear during the first few walking steps.
For the taped condition it was stretched to the maximum extent possible by hand and wrapped tightly (same experimenter for all subjects).
In the control condition, the tape was wrapped at the forefoot without perceivable stretch.
The foot was held unloaded at the time of application of the tape in the taped and control conditions.
Each trial was continuously monitored and repeated if the tape appeared to peel off or tear.
Steps where marker trajectories were missing or erroneous were rejected.

The added stiffness due to the elastic tape was measured and the repeatability of the experimenter's skill was assessed using dummy models in a materials testing machine (Instron model 3345, Norwood, MA).
The dummy forefoot comprised of two wooden blocks of 41.5\,mm length and 31\,mm width (\fig~\ref{fig:results:supplement:tape stiffness}a), whose dimensions are based on typical measurements of a human forefoot.
Five samples of the tape (3B Scientific, Hamburg, Germany) were tested by the same experimenter who performed the human subject trials.
The following protocol was used while measuring the force and displacement at 50~Hz:
\begin{enumerate}
  \item The blocks were rigidly clamped in a materials testing machine (Instron model 3345, Norwood, MA).
  \item A small initial separation between the blocks was chosen to allow a sheet of paper to easily slide between them (approximately 0.1\,mm).
  \item Tape of length 460\,mm was wrapped around the blocks after which the load cell and the displacement sensor were both zeroed.
  \item The block separation was cycled between 0 and 6\,mm using constant rate ramps of 0.16\,mm/s. Higher displacements were avoided as the tape started to peel away. 
  \item Based on pilot trials, four loading-unloading cycles were performed for each sample to allow the tape's elastic material to settle into a steady-state and the fifth loading cycle is reported. The initial displacement and load at 1\,mm extension in the fifth cycle are defined as zero (\fig~\ref{fig:results:supplement:tape stiffness}b).
\end{enumerate}

The stiffness, defined as the ratio of the total change in load to the total displacement, was 37.56\,N/mm$\pm$1.02\,N/mm (mean$\pm$standard deviation) in measurements from 5 different wrapping trials (\fig~\ref{fig:results:supplement:tape stiffness}b). 

\subsection{Data Analysis}
\label{sec:data analysis methods}

Kinematic and kinetic data were low-pass filtered using a zero-phase fourth order Butterworth filter with a cutoff frequency of 6\,Hz.
We use a multi-segment foot model comprising hindfoot and forefoot segments (\fig~\ref{fig:methods}b).
Segment definition for the hindfoot follows \citet{Leardini2007}.
Forefoot segment definition follows \citet{bruening2012analysis1} and relies on proximal markers on the first and fifth metatarsals and the distal second metatarsal marker.
This definition does not use the distal first and fifth metatarsal markers, which reduces two possible sources of variability in segment tracking arising from the taping intervention.
The distal first, second and fifth metatarsal markers were removed and replaced among the three taping conditions. 
By not relying on the distal first and fifth markers for segment definition, we minimize the between-condition differences in segment tracking caused by any potential errors in marker replacement at the forefoot.
Further, modes of deformation of the forefoot also affect our choice of segment definition.
The distance between the first and fifth metatarsal heads is known to change through stance as the forefoot splays transversely \citep{Leardini2007}, and forefoot taping directly affects the extent of splaying in our experiment (mean $\pm$ standard error of the mean (SEM): 5.6\,mm $\pm$ 0.3\,mm splay in the taped condition versus 9.4\,mm $\pm$ 0.6\,mm in the free condition).
The segment definition based on \citet{bruening2012analysis1} uses only the distal second metatarsal marker, along with proximal first and fifth metatarsal markers at the midfoot, thereby avoiding errors due to between-condition differences in forefoot deformation modes.

For inverse dynamics computations the midfoot joint is defined as the midpoint of the navicular marker and the base of the fifth metatarsal. 
Rotation between the hindfoot (parent) and forefoot (child) segments is defined using the following Euler angle scheme (1 : dorsiflexion -- plantarflexion, 2 : adduction -- abduction, and 3 : inversion -- eversion).
The corresponding rotation matrix $R$ was computed in \texttt{SciPy 1.7.0} and numerically differentiated to obtain the angular velocity of the forefoot in the hindfoot frame ($[\boldsymbol{\omega}]_\times = \dot{R}R^T$) and verified by comparison with the analytical expression for angular velocity for this rotation scheme,
\begin{align*}
  \omega_x &= \dot{\theta_3}\cos\theta_2\cos\theta_1 + \dot{\theta_2}\sin\theta_1 \\
  \omega_y &= \dot{\theta_1} + \dot{\theta_3}\sin\theta_2 \\
  \omega_z &= \dot{\theta_2}\cos\theta_1 - \dot{\theta_3}\cos\theta_2\sin\theta_1,
\end{align*}
where $\theta_1$, $\theta_2$, $\theta_3$ are the measured Euler angles and $\omega_x$,$\omega_y$ and $\omega_z$ are components of the angular velocity vector in the hindfoot frame.
Joint forces and torques at the midfoot were estimated using inverse dynamics in BodyLanguage (Vicon, Oxford, UK).
The joint rotational power was calculated as the dot product of the midfoot torque and the relative midfoot angular velocity ($P_{\rm rot} = \mathbf{M}\cdot\boldsymbol{\omega}$).
The joint translational power was calculated as the dot product of the midfoot reaction force and the relative midfoot velocity ($P_{\rm tra} = \mathbf{F}\cdot\mathbf{v}$).
Six DOF joint power at the midfoot joint is calculated as the sum of the rotational and translational power terms following \citep{buczek1994translational,takahashi2017energy}.
Negative dorsiflexion work during stance was calculated as the time-integral of the power curve when power was negative.
Work was calculated for each step in a trial and averaged over all steps to obtain the mean work for the trial.
Sagittal plane stiffness is calculated as the slope of the sagittal plane torque with respect to the sagittal plane deformation angle when the midfoot is dorsiflexing.
Sagittal plane torque is the mediolateral component of the midfoot torque expressed in the hindfoot coordinate frame.
Sagittal plane angle is the angle of rotation of the forefoot segment measured about the mediolateral axis of the fixed hindfoot frame.
Some subjects show a knee in the midfoot torque-angle curve when at lower midfoot torques.
So, linear regression is performed for the portion of the curve when midfoot torque is greater than one third of the maximum torque.

To quantify the effect of taping on transverse curvature at the midfoot, we measure the included angle between the line segments joining the first and second metatarsal bases, and the second and fifth metatarsal bases, projected onto the coronal plane of the heel frame.
This curvature quantifier uses anatomical landmarks within the foot to capture shape without being affected by overall rigid translations or rotations of the whole foot.
The angle at the forefoot is similarly measured using markers at the metatarsal heads.
In the forefoot as well as the midfoot, greater the transverse curvature, the smaller (more acute) the included angle.
Transverse angles are measured at the beginning of stance (heel contact) and at the instant when the midfoot torque is maximum during stance.
Forefoot width is defined as the distance between markers on the first and fifth metatarsal heads. 
Change in forefoot width, $\Delta w$, is the difference between forefoot width at maximum midfoot torque, and forefoot width at the beginning of stance.

Data were processed in \texttt{Python 3.8.1} using  \texttt{NumPy 1.21.1} and  \texttt{SciPy 1.7.0}, and in \texttt{MATLAB 9.8.0.1323502 (R2020a)}.

\subsection{Statistical methods} 
Between-group data independence is ensured through randomization of the experimental conditions. 
Homogeneity of variances is verified through Bartlett's test ($p > 0.2$ in all tests) and data normality is verified using the Shapiro-Wilk test ($p > 0.1$ in all tests).

We performed a Type-III one-way analysis of variance (ANOVA) with Satterthwaite's method to test the effect of applying the forefoot tape.
The ANOVAs were posed as a linear mixed model with condition (levels: free, tape, control) as the fixed factor and subject as random factor, given by
\begin{equation}
Y_{si} = \beta_0 + S_{0s} + \beta_iX_i + \epsilon_{si},
\end{equation}
where $s = 1 \to 13$ is the index for subjects, and $i = 1,2$ for taping condition. The dependent variable $Y$ has 39 values corresponding to 13 subjects and 3 conditions.
The global intercept $\beta_0$ (corresponding to the free condition) and the fixed effects $\beta_i$ (for control and tape conditions), are estimated.
The random effect of subjects $S_{0s}$ is assumed to be normally distributed with zero mean and encodes the between-subject variability in the intercept. 
The residual term $\epsilon_{si}$ is also assumed to be distributed normally.
Significance level for all statistical tests is set at 0.05.
If the ANOVA was significant, we performed a Tukey all-pairs comparison with Holm-Bonferroni correction.

Statistical tests were performed in R (v4.0.3) \citep{rLanguage}, the mixed model analysis was performed using the \texttt{lme4 v1.1.25}, ANOVA was performed using \texttt{lmerTest v3.1.3}, and post-hoc comparisons were performed using the \texttt{multcomp v1.4.15} package.

\section{Results}

\begin{figure}[bht!]
  \centering
  \includegraphics[width=0.5\textwidth]{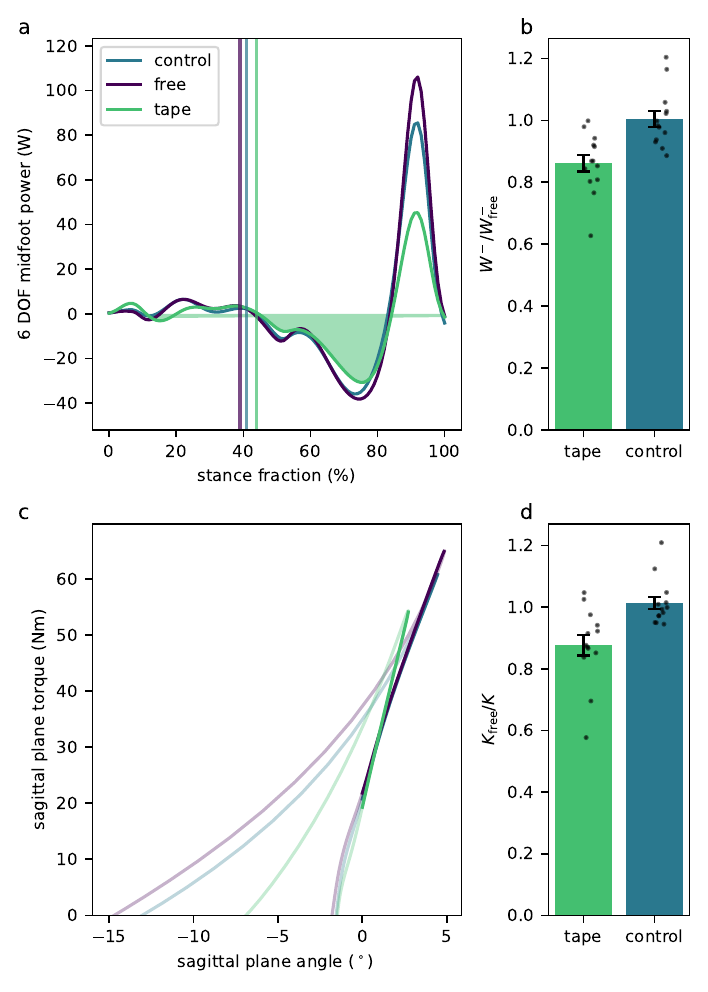}
  \caption[Change in midfoot work and stiffness during walking.]{{\bfseries Change in midfoot work and stiffness during walking.}
  {\bfseries a,} Six degrees of freedom midfoot power through stance in the free, taped and control conditions. The net negative work done during stance in the taped condition (shaded region) is smaller than in the free or control conditions. Vertical lines show the instant in stance when the center of pressure crossed the midfoot joint. 
  {\bfseries b,} Negative work over stance, normalized by the free condition $(W/W_{\rm free})$ and averaged across all subjects (n = 13).
  Whiskers show the SEM.
  {\bfseries c,} Sagittal torque at the midfoot joint versus the midfoot deformation angle. Thick lines show the portion of the load-deformation curve used for midfoot stiffness estimation.
  {\bfseries d,} Midfoot stiffness in the free condition relative to the taped and control conditions $(K_{\rm free}/K)$ (n = 13).
  Whiskers show the SEM.
  Correlation between work and stiffness is reported in \fig~\ref{fig:results:supplement:correlation}.
  Data on positive work production and toe angle change between conditions are reported in \fig~\ref{fig:results:supplement:push off} and \fig~\ref{fig:results:supplement:toe}.
  Data on forefoot width variation through stance are reported in \fig~\ref{fig:results:supplement:splay}.
  Correlations between forefoot curvature and midfoot work and stiffness are reported in \fig~\ref{fig:results:supplement:curvature}.
  Individual subject power data are reported in \fig~\ref{fig:results:supplement:power traces},
  and torque-angle curves in \fig~\ref{fig:results:supplement:moment angle traces}.
  Raw data underlying this figure and its supplements are available in figure 3 -- source data 1--7.
  }\label{fig:results}
\end{figure}

Dorsiflexion (negative) power at the midfoot is smaller through stance for the taped condition compared to the free or control conditions (representative data in \fig~\ref{fig:results}a).
Consequently, relative to the free condition, the negative work at the midfoot is smaller in the taped condition.
Taping condition is a significant factor for the ANOVA on midfoot negative work ($F_{2,24} = 12.09, p = 0.0002$).
The taped and free conditions are significantly different from each other ($z = 4.42, p < 0.0001$) while the control condition is not statistically different from the free condition ($z = 0.34, p > 0.999$).
On average, the negative work for the taped condition is lower than the free condition by $13.9\%\pm3\%$ ((mean\,$\pm$\,SEM), \fig~\ref{fig:results}b).

In addition to work, we also analyzed midfoot stiffness, the slope of the sagittal plane midfoot torque with respect to sagittal plane midfoot angle (\fig~\ref{fig:results}c), as an independent test of the hypothesis that midfoot stiffness was altered by the tape.
Taping condition is a significant factor for the ANOVA on midfoot stiffness ($F_{2,24} = 9.57, p = 0.0009$).
Midfoot stiffness is significantly higher for the taped condition compared to the free condition by $16.8\% \pm 5.8\%$ ((mean\,$\pm$\,SEM), $z = 3.66, p = 0.0007$,\fig~\ref{fig:results}d), while the control condition is not distinguishable from the free condition ($z = -0.24, p > 0.999$).
These stiffness changes agree with and corroborate the changes in midfoot work with taping.
Indeed, linear regression shows that the work and stiffness ratios for the taped foot relative to the free foot are significantly correlated ($R^2 = 0.64$, $p = 0.001$; \fig~\ref{fig:results:supplement:correlation}).
Therefore, we conclude that transversely taping the forefoot significantly reduces the negative work at the midfoot and increases the midfoot joint stiffness during walking (\fig~\ref{fig:results}).

Changes were observed in the peak power and the positive work output of the midfoot, whose implications to the windlass mechanism are discussed in section~\ref{section:speculation}.
Peak positive power in the taped and control conditions are significantly smaller than the free condition ($z = -6.94, p < 0.0001$, and $z = -2.45, p = 0.04$ respectively, \fig~\ref{fig:results:supplement:push off}a).
Peak negative power is not statistically distinguishable among conditions ($F_{2,24} = 2.99, p = 0.07$, \fig~\ref{fig:results:supplement:push off}b).
Positive midfoot work is significantly different between the taped and free conditions ($z = -5.3, p < 0.0001$), but the control condition is not statistically different from the free condition ($z = -1.08, p = 0.8$, \fig~\ref{fig:results:supplement:push off}c).

To test the idea that changes in the positive work are related to the windlass and not TTA function, we analyzed how toe dorsiflexion was affected by the taping condition and its correlation with work.
The extent of toe dorsiflexion has no correlation with the negative midfoot work ($R^2 < 0.05$ for each condition; \fig~\ref{fig:results:supplement:toe}a), but it is correlated with positive midfoot work in each condition (free: $R^2 = 0.52, p = 0.006$; control: $R^2 = 0.50, p = 0.007$; tape: $R^2 = 0.76, p < 0.0001$; \fig~\ref{fig:results:supplement:toe}b).
Extent of toe dorsiflexion in the taped condition is significantly smaller than the free condition ($z=-5.56, p < 0.0001$), but the control condition is indistinguishable from the free condition ($z=-2.24, p = 0.08$, \fig~\ref{fig:results:supplement:toe}c).

Taping affected forefoot geometry but not the midfoot arch.
Forefoot width is smaller in the taped condition than the free or control conditions (\fig~\ref{fig:results:supplement:splay}a).
Change in forefoot width normalized by the root mean squared midfoot torque is smaller in the taped condition relative to the free condition by about 35\% ($z=-7.66, p < 0.0001$), but the control and free conditions are not statistically distinguishable ($z=-1.40, p=0.48$, \fig~\ref{fig:results:supplement:splay}b).
To evaluate if the observed changes in midfoot work are related to the extent of forefoot splay restriction, we measured correlation between negative midfoot work and forefoot splay restriction, which is defined as the change in forefoot width from the beginning of stance to the point of maximum midfoot torque.  
We found a positive correlation between midfoot work and forefoot splay restriction ($R^2 = 0.51, p =0.006$, \fig~\ref{fig:results:supplement:splay}c).

There was no statistically significant change in the midfoot TTA curvature irrespective of the load on the midfoot (beginning of stance: $F_{2,24}=0.72, p = 0.49$; when the midfoot torque is maximum: $F_{2,24}=0.37,p=0.69$).
However, taping increased the forefoot transverse curvature at all foot loads (beginning of stance: $F_{2,24} =38.21, p < 0.0001$; at maximum midfoot torque: $F_{2,24} = 11.74, p=0.0003$).
There is no significant correlation between forefoot curvature of the taped foot and change in midfoot negative work, either at the beginning of stance ($R^2 = 0.09, p = 0.32$) or at maximum midfoot torque ($R^2 = 0.01, p = 0.71$, \fig~\ref{fig:results:supplement:curvature}a).
Similarly, there is no correlation between forefoot curvature and midfoot stiffness (beginning of stance: $R^2 < 0.01, p = 0.84$; at maximum midfoot torque: $R^2 < 0.01, p = 0.98$, \fig~\ref{fig:results:supplement:curvature}b).


\section{Discussion}

We have shown that the cross-axis coupling found in cadaveric studies is also effective during walking despite the likely engagement of the foot's muscles for propulsion.
But this should not be interpreted to mean that muscles cannot or do not modulate the TTA's contribution.
The tibialis posterior, which is known to affect MLA deformation \citep{johnson1989tibialis,kitaoka1997effect,kohls2004tibialis}, also inserts into the tarsal bones and into multiple proximal metatarsal areas that are part of the TTA's structure.
Indeed, simulated tibialis posterior contraction has been previously reported to alter the TTA shape in cadaveric tests \citep{kitaoka1997effect}.
The intrinsic dorsal interossei and adductor hallucis muscles could also be important because they have transverse lines of action \citep{kura1997quantitative}, and their contraction could make use of the TTA-induced cross-axis coupling by altering the intermetatarsal stiffness.
Nevertheless, our measurements show that the cross-axis coupling that is seen in cadaveric feet carries over to walking.

Between-subject variability in our data may arise from several factors, including the tape wrapping method's repeatability, and the tape application location relative to the effective midfoot bending axis.
One possible issue, is the effect of taping on gait parameters.
To minimize that, subjects walked at the same normalized speed between-conditions to minimize the effect of speed on foot stiffness that has been observed during locomotion \citep{holowka2020human,caravaggi2010dynamics}.
Additionally, taping did not affect the stance phase duration ($F_{2,24}=1.18, p=0.32$), and there were no observed differences in center of pressure trajectories (\fig~\ref{fig:results:supplement:COP medio-lateral traces}, \fig~\ref{fig:results:supplement:COP antero-posterior traces}), or ankle angle during stance (\fig~\ref{fig:results:supplement:ankle angle traces}).
The tape was applied manually by the experimenter, and so we took additional steps to assess the repeatability of the tape wrapping protocol (\fig~\ref{fig:results:supplement:tape stiffness}).

Mathematical analysis of a previously derived model for a mechanical foot-like device reveals other sources of variability that are not easily controlled.
The longitudinal bending stiffness depends on the square of the anteroposterior distance between the metatarsal hinges and the transverse springs at the distal end \citep[equations (S4.3) and (S4.4) in][]{venkadesan2020stiffness}.
Therefore, small variability in the anteroposterior placement of the transverse springs relative to the hinges will be amplified at least two-fold when it maps to variability of the longitudinal bending stiffness.
We expect a similar issue with the anteroposterior tape location.
Although the experimenter used clear bony landmarks of the distal metatarsal heads to wrap the tape, the quantity that matters is the distance between the tape location and the effective axis of rotation at the midfoot.
There is unknown variability in the anatomical equivalent of the midfoot hinge because the midfoot is comprised of a multi-articular arrangement of bones that is not a simple hinge-like joint.
The effective midfoot center of rotation may lie somewhere between the tarsometatarsal joints and the subtalar joint, although it is probably close to the mid-dorsum area \citep{ito2017three}.
The midfoot articular region occupies a substantial fraction of the foot's length and thus natural anatomical variability may lead to considerable variability in how the elastic tape's stiffness is transformed into longitudinal midfoot bending stiffness.
Other anatomically variable factors include forefoot splay, TTA and MLA curvatures, foot lever length, and tissue stiffness which would affect the tape's stiffness relative to each individual subject's intermetatarsal stiffness.
Refinements to reduce variability, such as modifying the device used to increase transverse stiffness and subject-specific measurements of midfoot geometry and intermetatarsal stiffness are left for future studies.

Recent work has attributed the increase in foot stiffness upon applying a transverse forefoot tape to an increase in midfoot and forefoot transverse curvatures caused by the taping \citep{tang2025wrapping}.
By contrast, our data demonstrate that the TTA can stiffen the foot without changes in the arch geometry.
We found that taping increased foot stiffness despite no statistically significant change in the midfoot TTA curvature either at the beginning of stance ($F_{2,24}=0.62, p = 0.55$) or when the midfoot torque was maximum ($F_{2,24}=0.41,p=0.67$).

Despite no change in TTA curvature, the foot stiffens due an increase in intermetatarsal stiffness upon taping, analogous to the behavior of the foot model in \fig~\ref{fig:intro}c.
While taping did increase the transverse curvature of the forefoot, this had no measurable effect on midfoot work or stiffness. 
Forefoot curvature of the taped foot had no statistically significant correlation with change in foot stiffness or midfoot work caused by taping (\fig~\ref{fig:results:supplement:curvature}).
Therefore, in contrast with \citet{tang2025wrapping}, we found that taping increased foot stiffness without altering TTA curvature, indicating that the mechanism is stiffness transmission rather than shape change, consistent with the cross-axis coupling hypothesis \citep{venkadesan2020stiffness}.
This suggests that foot shape-based explanations alone cannot account for the observed changes, and that the contribution of transverse inter-metatarsal stiffness to overall midfoot bending stiffness is the main factor.



\subsection{Ideas and speculation}
\label{section:speculation}
In addition to reduced negative midfoot work, forefoot taping also decreased the peak positive push-off power and the late stance positive midfoot work (\fig~\ref{fig:results:supplement:push off}a,c).
We propose that these reductions are not due to increased foot stiffness upon taping, but are instead because the tape affected toe dorsiflexion, which in turn affects windlass contributions to midfoot work \citep{yawarStructuralOriginsStiffness2022,welte2018influence}.
The transverse tape at the forefoot resulted in reduced toe dorsiflexion during push-off (\fig~\ref{fig:results:supplement:toe}c), which we speculate as reducing the level of windlass engagement and thus reducing the midfoot positive work in late stance (\fig~\ref{fig:results:supplement:push off}c).
This speculation is supported by the observed correlation of toe dorsiflexion with midfoot positive work for every condition (\fig~\ref{fig:results:supplement:toe}b).
Whereas, no such correlation is found between toe dorsiflexion and midfoot negative work (\fig~\ref{fig:results:supplement:toe}a).
A detailed investigation of the windlass mechanism's role in the positive push-off work is the topic of a separate study, but our data support the thinking that toe dorsiflexion during push-off affects the late stance midfoot work due to its role in windlass function and not because it alters the TTA's function \citep{yawarStructuralOriginsStiffness2022}.

There are several potential applications of our work in podiatry and sports biomechanics.
Adult acquired flatfoot disorders \citep{kohls2004tibialis}, especially in patients with advanced diabetes \citep{hastings2014kinematics,gelber2014windlass}, could cause significant impairment to locomotion ability due to increased foot flexibility.
Future clinical studies are needed but non-invasive foot taping could serve as a strategy to manage these conditions.
Foot taping in therapeutic settings reduces MLA deformation during quiet standing or walking \citep{ator1991effect,yoho2012biomechanical}, even for people with flat longitudinal arches \citep{bishop2016effects}. 
Unlike our taping method, these studies applied tape all over the foot, including the midfoot and proximal regions around the heel.
In patients with vascular issues, such as the case with diabetes, reducing the amount of compressive taping may be desirable.
Our results show that increasing transverse stiffness in the bony regions at the ball of the foot is sufficient, which could be better suited for patients with vascular issues than taping the whole foot.
Similarly, athletic performance in sports may depend on foot stiffness \citep{Brinkmann2020aa}, and a stiff distal tape may augment midfoot stiffness without adding much weight.




\section*{Acknowledgments}
Volunteers for participation, and Nick Bernardo, Dylan Shah, and the Yale Center for Engineering Innovation and Design for fabrication and mechanical testing support.

\section*{Competing interests}
The authors declare no competing financial interests.

\section*{Funding}
This material is based upon work supported by the Human Frontier Science Program and by the National Science Foundation under Grant Number 2046120.

\section*{Author contributions}
MV and SM conceived of the study. AY and LK conducted the human subject experiments with guidance from MV.\@ AY analyzed the data and performed statistics in consultation with MV.\@ AY and MV wrote the manuscript, and all authors contributed edits.

\singlespacing{}
\nolinenumbers{}
\bibliographystyle{jxb} 

\clearpage

\setcounter{figure}{0}




\section*{Figure supplements}
\setcounter{figure}{0}
\renewcommand{\thefigure}{\ref{fig:methods} -- figure~supplement~\arabic{figure}}

\begin{figure}[bht!]
  \centering
  \includegraphics[width=0.6\textwidth]{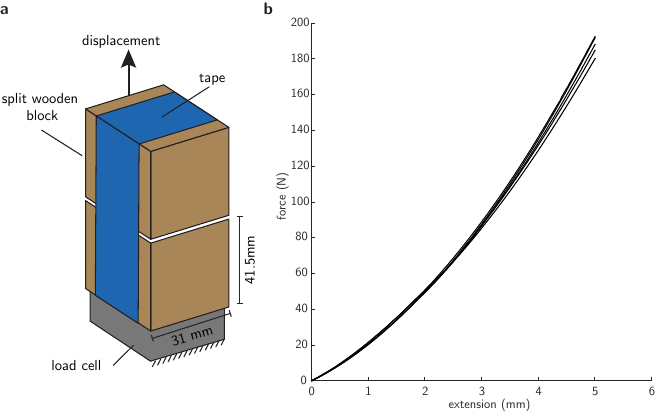}
  \caption[Assessment of the tape wrapping protocol.]{
    {\bfseries Assessment of the tape wrapping protocol.} {\bfseries a,} Elastic tape was wrapped around two wooden blocks rigidly clamped to a materials testing machine. The blocks were cyclically separated to measure the stiffness of the wrapped tape.
  {\bfseries b,} Force-displacement curves of five samples of the tape were measured to assess the variability in the wrapping protocol.
  The fifth cycle for each sample is shown here.
  From 5 repetitions of the wrapping and testing steps, the ratio of the total change in load to the total displacement, or the effective stiffness, is $37.56\,{\rm N/mm} \pm 1.02$\,N/mm  (mean\,$\pm$\,standard deviation).
  \label{fig:results:supplement:tape stiffness}}
\end{figure}

\setcounter{figure}{0}
\renewcommand{\thefigure}{\ref{fig:results} -- figure~supplement~\arabic{figure}}

\begin{figure}[bht!]
  \centering
  \includegraphics[width=0.5\textwidth]{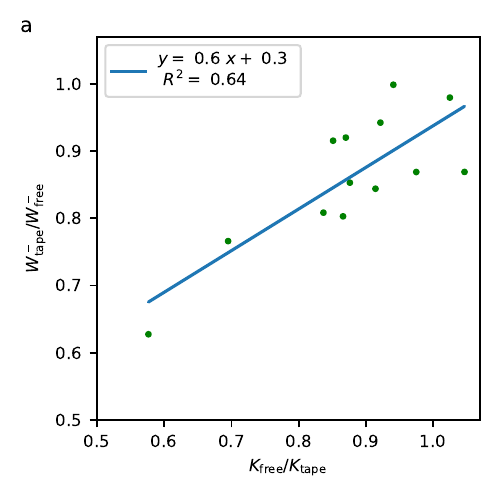}
  \caption{{Correlation between the ratio of negative midfoot work in the tape and free conditions, and ratio of midfoot stiffness in the free and tape conditions.}
  }\label{fig:results:supplement:correlation}
\end{figure}

\begin{figure}[bht!]
  \centering
  \includegraphics[width=\textwidth]{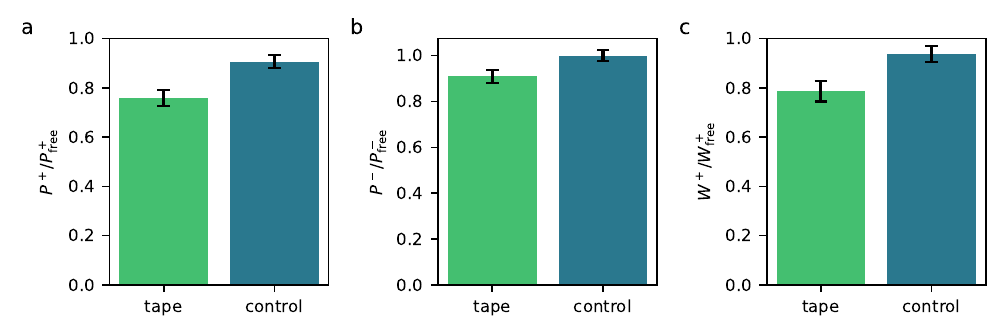}
  \caption{\textbf{Effect of experimental conditions on midfoot positive power and work.} Bar plots of {\bfseries a,} peak positive power and {\bfseries b,} peak negative power at the midfoot, normalized by the free condition. Bar plots of the {\bfseries c,} positive midfoot work over stance, normalized by the free condition. Whiskers show the SEM ($n = 13$).}
  \label{fig:results:supplement:push off}
\end{figure}

\begin{figure}[bht!]
  \centering
  \includegraphics[width=\textwidth]{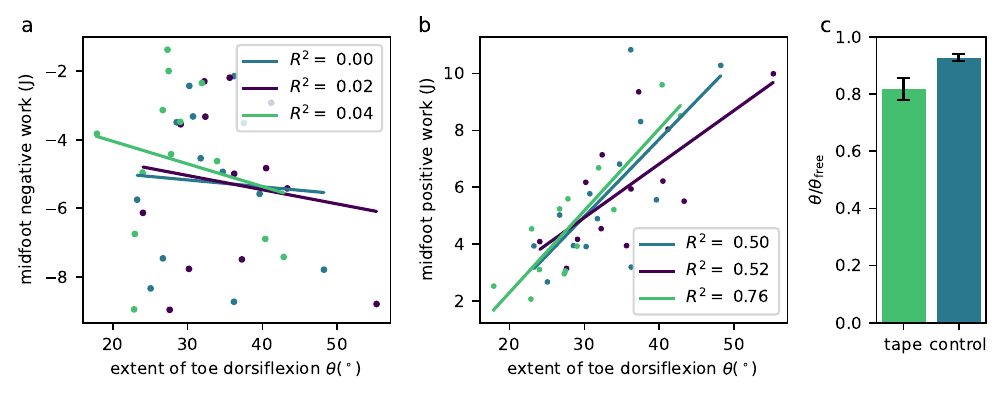}
  \caption{\textbf{Variation in midfoot work with extent of toe dorsiflexion.}  {\bfseries a,} midfoot negative work and {\bfseries b,} midfoot positive work over stance versus toe dorsiflexion ($\theta$) at push-off in all three experimental conditions. Each dot represents one participant. Solid lines are linear regression fits. {\bfseries c,} Bar plots of the extent of toe dorsiflexion ($\theta$) relative to the free condition. Whiskers show the SEM ($n = 13$).}
  \label{fig:results:supplement:toe}
\end{figure}

\begin{figure}[bht!]
  \centering
  \includegraphics[width=\textwidth]{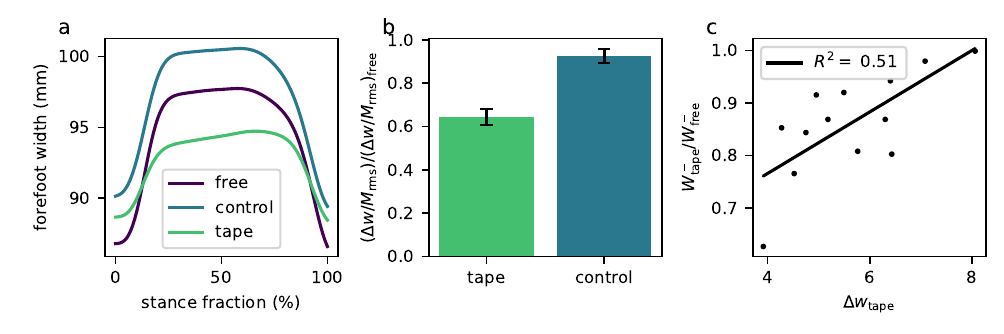}
  \caption{{\bfseries Measured change in forefoot width during walking.}
  {\bfseries a,} Traces of forefoot width $w$ through stance in each condition, averaged over all participants ($n = 13$). Change in forefoot width $\Delta w$ is defined as the difference between forefoot width at maximum midfoot torque and forefoot width at the beginning of stance.
  {\bfseries b,} Bar plots of change in forefoot width ($\Delta w$) divided by the root mean squared midfoot torque ($M_{\rm rms}$) for the tape and control conditions relative to the free condition. Whiskers show the SEM ($n = 13$). 
  {\bfseries c,} Change in forefoot width in the taped condition over stance $\Delta w_{\rm tape}$  versus the ratio of negative midfoot work in the taped and free conditions. Each dot corresponds to a participant.
  }\label{fig:results:supplement:splay}
\end{figure}

\begin{figure}[bht!]
  \centering
  \includegraphics[width=0.8\textwidth]{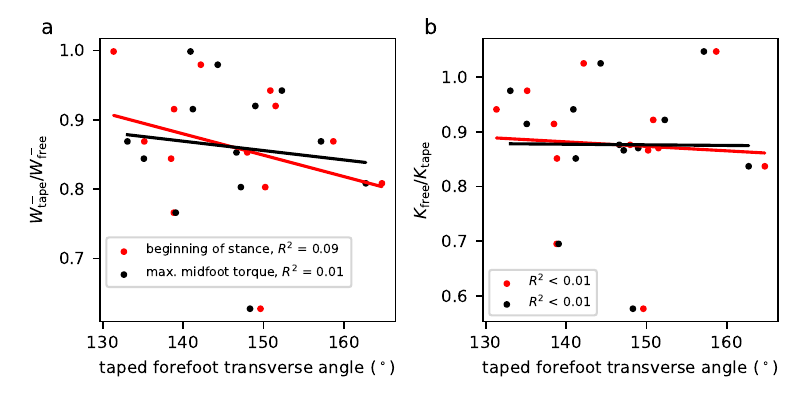}
  \caption{{\bfseries Effect of forefoot transverse angle on negative midfoot work and midfoot stiffness.}
  Greater the angle, smaller the curvature.
  {\bfseries a,} The ratio of negative midfoot work in the tape and free conditions versus the forefoot transverse angle at the beginning of stance (red) and when the midfoot torque is maximum (black).
  {\bfseries b,} The ratio of midfoot stiffness in the free and tape conditions versus the forefoot transverse angle at the beginning of stance (red) and when the midfoot torque is maximum (black). Each dot corresponds to a participant, and solid lines are linear regression fits.
  }\label{fig:results:supplement:curvature}
\end{figure}

\begin{figure}[bht!]
  \centering
  \includegraphics[width=0.8\textwidth]{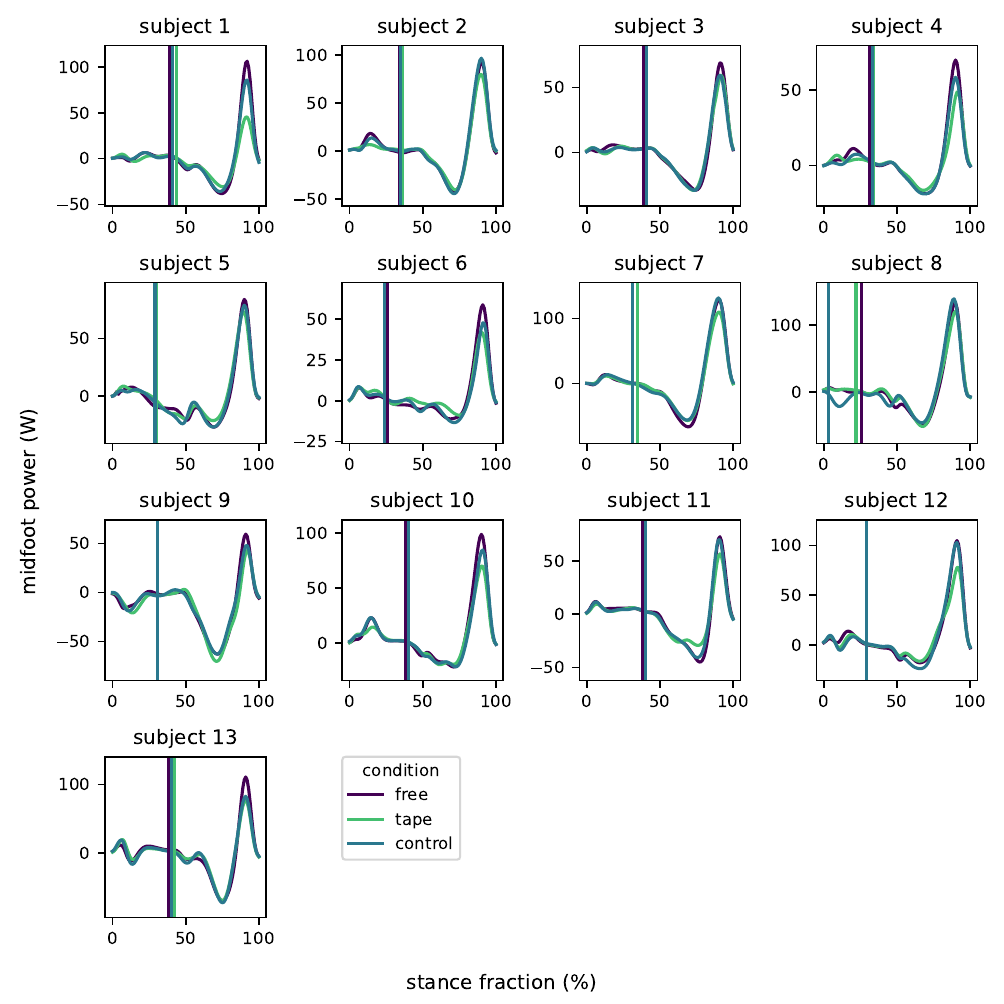}
  \caption{Midfoot power through stance for all subjects in all conditions. Vertical lines show the instant of time when the center of pressure crossed the midfoot joint.}
  \label{fig:results:supplement:power traces}
\end{figure}

\begin{figure}[bht!]
  \centering
  \includegraphics[width=0.8\textwidth]{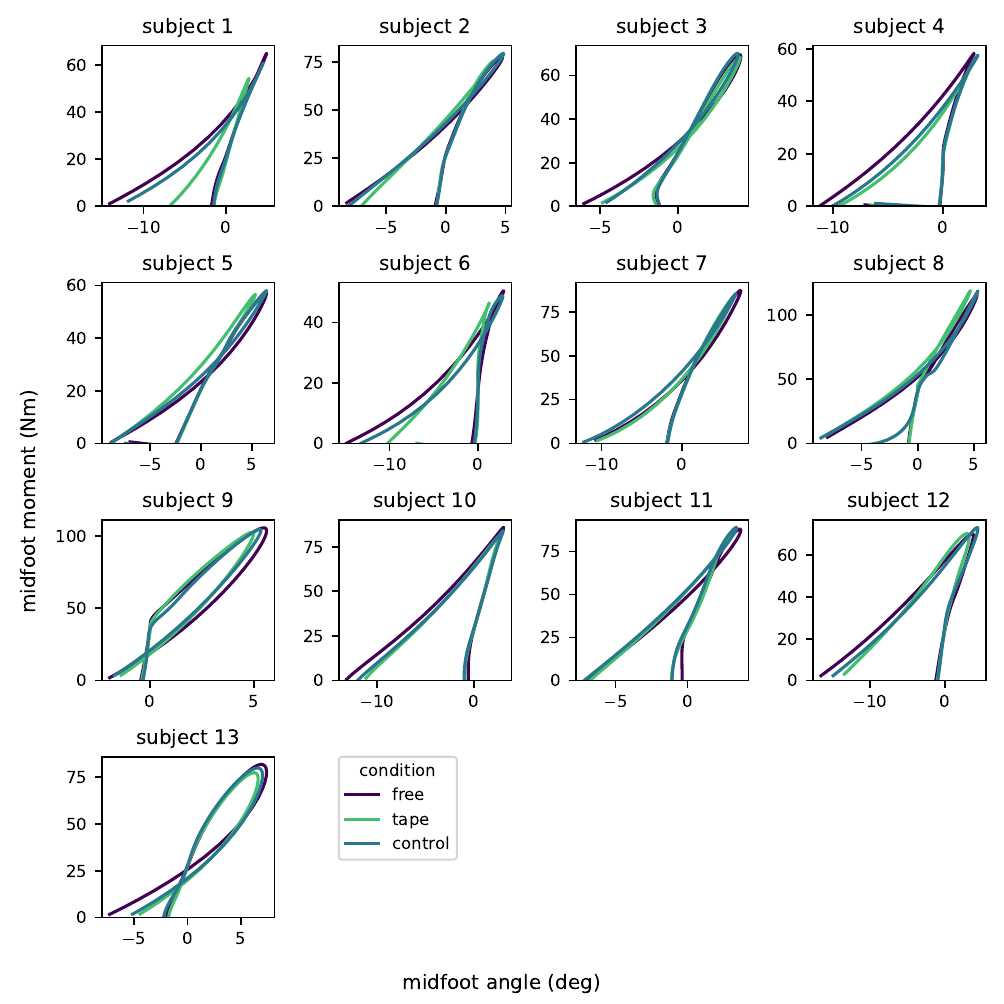}
  \caption[Sagittal midfoot torque vs sagittal midfoot angle for all subjects in all conditions.]{Sagittal midfoot torque vs sagittal midfoot angle for all subjects in all conditions.}\label{fig:results:supplement:moment angle traces}
\end{figure}

\begin{figure}[bht!]
  \centering
  \includegraphics[width=0.8\textwidth]{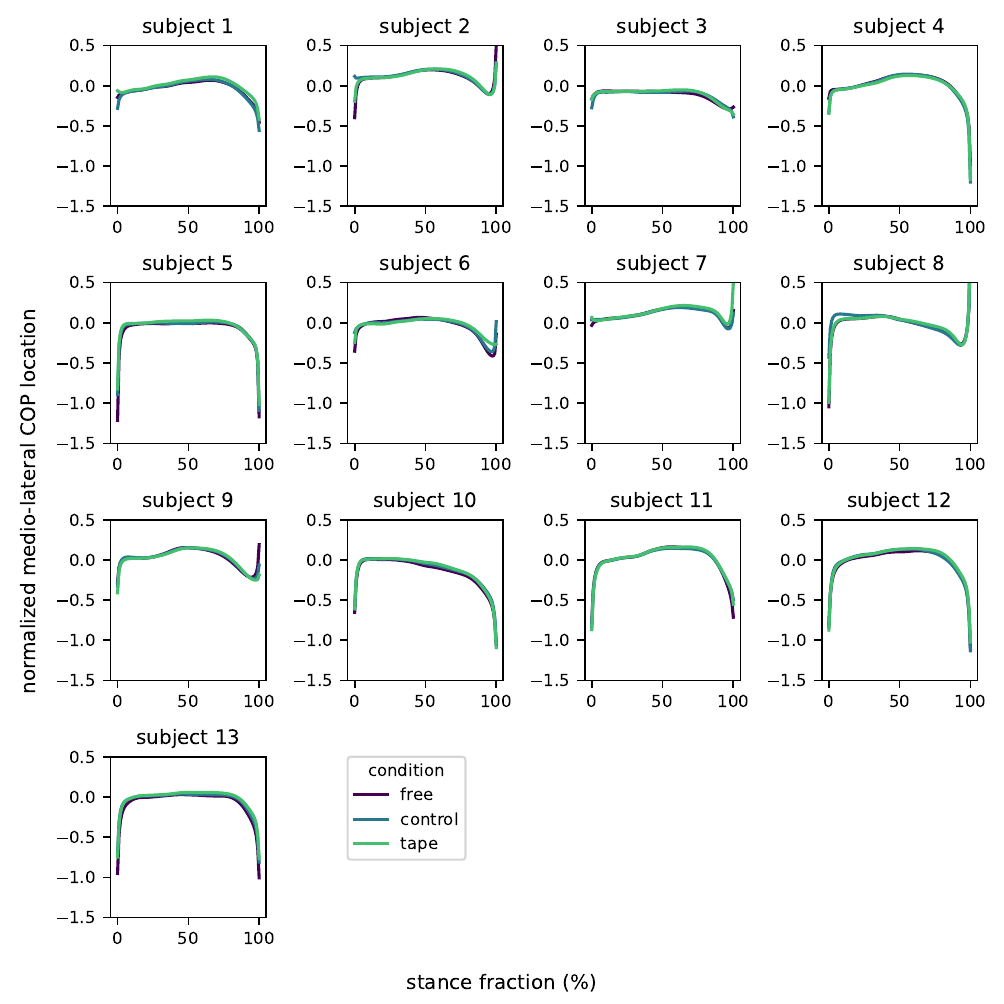}
  \caption[Mediolateral location of the center of pressure in the hindfoot reference frame through stance for all subjects in all conditions.]{Mediolateral location of the center of pressure in the hindfoot reference frame through stance for all subjects in all conditions. The center of pressure location is normalized by forefoot width.}\label{fig:results:supplement:COP medio-lateral traces}
\end{figure}

\begin{figure}[bht!]
  \centering
  \includegraphics[width=0.8\textwidth]{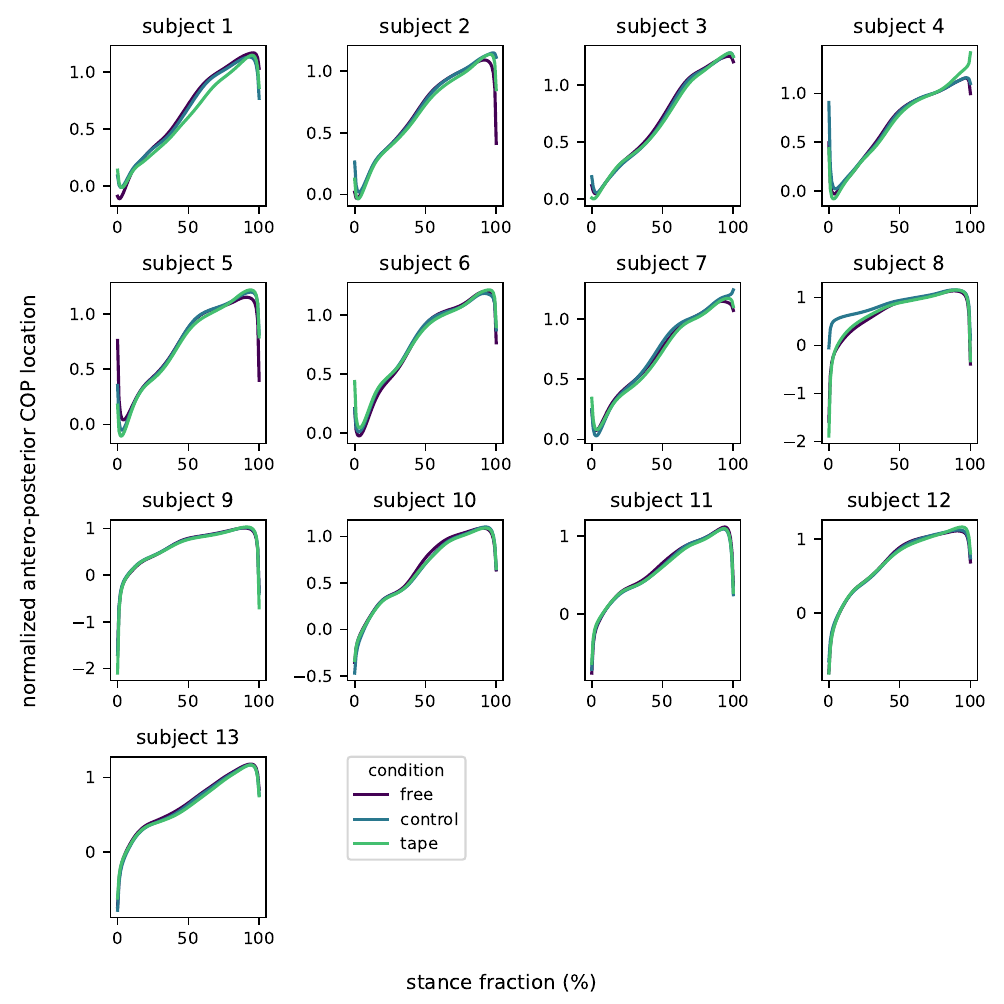}
  \caption[Anteroposterior location of the center of pressure in the hindfoot reference frame through stance for all subjects in all conditions.]{Anteroposterior location of the center of pressure in the hindfoot reference frame through stance for all subjects in all conditions. The center of pressure location is normalized by truncated foot length.}\label{fig:results:supplement:COP antero-posterior traces}
\end{figure}

\begin{figure}[bht!]
  \centering
  \includegraphics[width=0.8\textwidth]{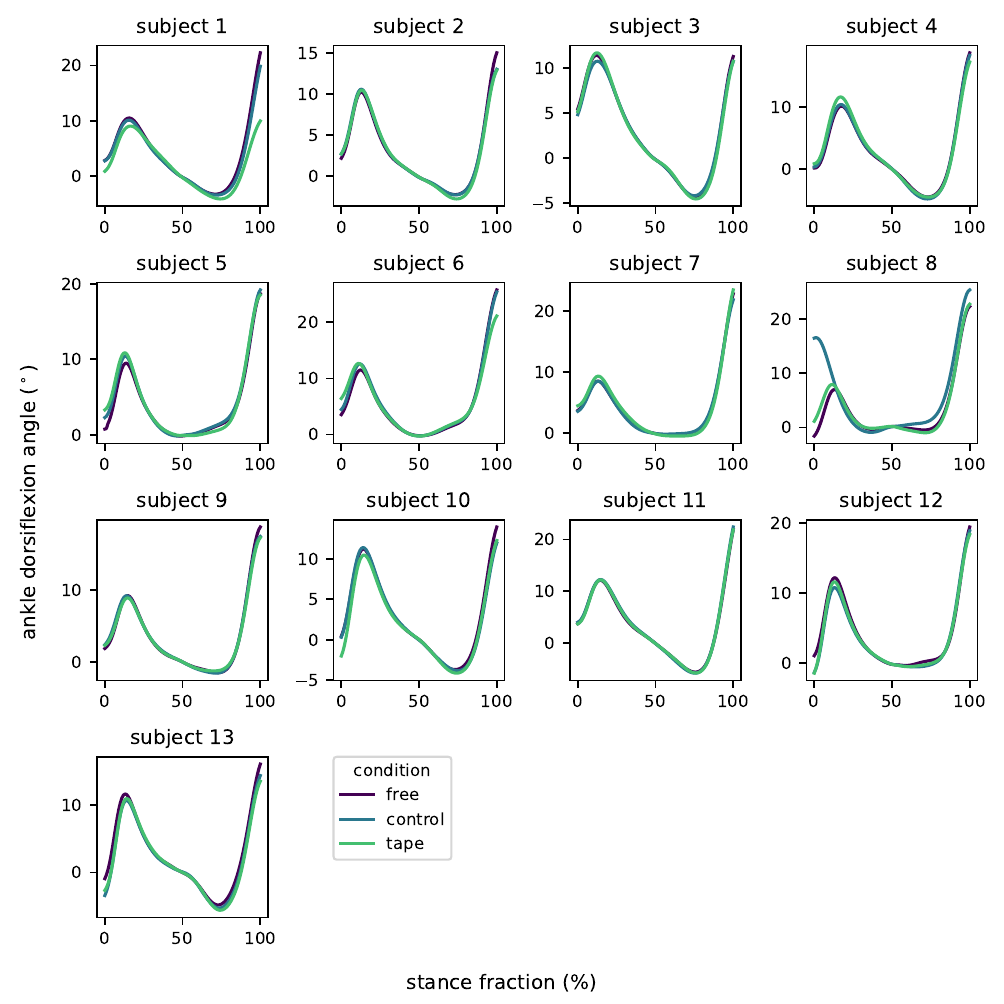}
  \caption[Ankle dorsiflexion angle through stance for all subjects in all conditions.]{Ankle dorsiflexion angle through stance for all subjects in all conditions. Zero dorsiflexion corresponds to foot-flat and a vertical shank.}\label{fig:results:supplement:ankle angle traces}
\end{figure}

\begin{table}[bht!]
\centering
\begin{tabular*}{0.6\textwidth}{c @{\extracolsep{\fill}} cc}
  \hline
  measurement & median & interquartile range \\
  \hline
 distance 1 & 239.46\,mm & 0.27\,mm  \\
 distance 2 & 160.27\,mm & 0.13\,mm \\
 distance 3 & 79.71\,mm & 0.25\,mm \\
 distance 4 & 119.83\,mm & 0.19\,mm \\
 right angle 1 & 89.64\,$^\circ$ & 0.15\,$^\circ$\\
 right angle 2 & 89.69\,$^\circ$ & 0.13\,$^\circ$\\
 right angle 3 & 89.53\,$^\circ$ & 0.13\,$^\circ$ \\
 \hline
\end{tabular*}
\caption[Motion capture calibration data]{\textbf{Motion capture calibration data} Uncertainty in distances between pairs of markers, and right angles between triplets of markers on a standard calibration object in a dynamic calibration trial. Median and interquartile range are reported as these measurements were not normally distributed.}
\label{table:methods:supplement:calibration}
\end{table}

  

\end{document}